\documentclass[a4paper,11pt]{article}
\usepackage{jheppub}
\usepackage[colorlinks=true,linkcolor=blue,citecolor=blue,urlcolor=blue]{hyperref}
\usepackage{amsfonts,amsmath,amssymb,mathrsfs}
\usepackage{mathtools,mathptmx,array}
\usepackage{subfigure,graphicx}

\def\bea{\begin{eqnarray}}
\def\eea{\end{eqnarray}}
\def\be{\begin{equation}}
\def\ee{\end{equation}}
\def\nn{\nonumber}
\def\p{\partial}

\def\cN{\mathcal{N}}
\def\cO{\mathcal{O}}

\title{Ultra-spinning Chow's black holes in six-dimensional gauged supergravity and their properties}

\author{Di Wu, Shuang-Qing Wu}

\affiliation{College of Physics and Space Science, China West Normal University, 1 Shida Road, Nanchong,
Sichuan 637002, People's Republic of China}

\emailAdd{wdcwnu@163.com}
\emailAdd{sqwu@cwnu.edu.cn}

\abstract{By taking the ultra-spinning limit as a simple solution-generating trick, a novel class
of ultra-spinning charged black hole solutions has been constructed from Chow's rotating charged
black hole with two equal-charge parameters in six-dimensional $\cN = 4$ gauged supergravity theory.
We investigate their thermodynamical properties and then demonstrate that all thermodynamical
quantities completely obey both the differential first law and the Bekenstein-Smarr mass formula.
For the six-dimensional ultra-spinning Chow's black hole with only one rotation parameter, we show
that it does not always obey the reverse isoperimetric inequality, thus it can be either sub-entropic
or super-entropic, depending upon the ranges of the mass parameter and especially the charge parameter.
This property is obviously different from that of the six-dimensional singly-rotating Kerr-AdS
super-entropic black hole, which always strictly violates the RII. For the six-dimensional
doubly-rotating Chow's black hole but ultra-spinning only along one spatial axis, we point
out that it may also obey or violate the RII, and can be either super-entropic or sub-entropic
in general.}


\begin{document}

\maketitle

\flushbottom

\section{Introduction}

Black holes in general relativity are some of the most remarkable and quite fascinating objects
in nature. Therefore, constructing exact solutions that present novel characteristics contribute
to our understanding of the general properties of black holes. A fundamental result in black
hole researches is Hawking's theorem about the horizon topology of the black holes. Notably, he
\cite{CMP25-152} had shown that the event horizons of four-dimensional stationary, asymptotically
flat black holes satisfying the dominant energy condition necessarily must be topology of $S^2$-sphere.
More interesting black objects can exist in four and higher dimensions if some of the assumptions
initially made in Hawking's theorem are relaxed. For instance, the uniqueness theorem that was
well-established for the four-dimensional black holes has been shown obviously no longer to hold
true when it is extended to five dimensions, and a lot of spacetimes in higher dimensions can
have several different kinds of horizon topologies as well as different asymptotical topological
structures at infinity. While both being asymptotically flat in five dimensions, a black ring
solution that owns a non-spherical $S^2 \times S^1$ horizon topology was first found in ref.
\cite{PRL88-101101}, apart from the known Myers-Perry black hole solution \cite{AP172-304}
that has a common horizon topology of the $S^3$-sphere. Later, a rotating black lens
solution which has the horizon topology of a lens-space $L(n,1)$ in the five-dimensional
asymptotically flat spacetime was also constructed in ref. \cite{PRD78-064062} (see recent
refs. \cite{KL2014,KL2016,TN2016,TO2017,ST2018a,ST2018b,ST2019} for more new black lens
solutions). In addition to these, by simply adding a spatial-like dimension to the four-dimensional
asymptotically flat black hole, one can obtain accordingly a trivial black string solution in
five dimensions. Another possibility is to relax asymptotic flatness. For example, in four-dimensions
with a negative cosmological constant, the Einstein equations admit topological black hole
solutions with the horizons being Riemann surfaces of any genus-$g$ \cite{PRD56-6475,CQG14-L109,
PLB353-46,PRD54-4891,AIPS13-311}. More generally, event horizons which are Einstein manifolds of
positive, zero, or negative curvature are possible in $D$-dimensional asymptotically anti-de Sitter
(AdS) space \cite{CQG14-L109,CQG16-1197}.

Recently, a new class of so-called ultra-spinning black holes \cite{PRL115-031101,PRD89-084007,
JHEP0114127,JHEP0220195} in which one of their rotation angular velocities is boosted to the
speed of light, has received considerable interest and enthusiasm. This kind of black hole
solutions which can exist in arbitrary dimension $D\ge 4$ have a noncompact horizon topology
but with a finite area, and are asymptotically (locally) AdS with horizons being topologically
spheres with two punctures at the north and south poles. The reverse isoperimetric inequality
(RII) \cite{PRD84-024037,PRD87-104017} indicates that the Schwarzschild-AdS black hole has the
maximum upper entropy, while the area entropy of this kind of black holes exceeds the maximum
entropy limit. Thus, it provides the first counter-example to this inequality, and such a black
hole is dubbed as ``super-entropic". Up to date, a lot of new ultra-spinning black hole solutions
\cite{JHEP0615096,JHEP0816148,PRD95-046002,1702.03448,JHEP0118042,PRD102-044007,PRD103-044014}
have been obtained from the known common rotating AdS black holes. In addition, various aspects
of the ultra-spinning black holes, including thermodynamical properties \cite{PRL115-031101,
JHEP0615096,PRD95-046002,1702.03448,JHEP0118042,MPLA35-2050098,PRD101-086006,PRD101-024057,
PLB807-135529}, horizon geometry \cite{PRD89-084007,JHEP0615096,PRD95-046002}, Kerr/CFT
correspondence \cite{PRD95-046002,1702.03448,JHEP0816148}, geodesic motion \cite{JMP61-122504}
and null hypersurface caustics \cite{CQG38-045018,PRD103-024053,PRD103-104020}, etc, have
been widely explored.

Although there has been much progress in constructing ultra-spinning black hole solutions and
studying their physical properties, there are relatively fewer ultra-spinning black hole solutions
so far obtained in higher dimensional AdS spacetimes, and there is also a gap in the research
on their thermodynamical properties. Thus, many other ultra-spinning black hole solutions,
especially those in higher dimensional gauged supergravity theories, and their thermodynamical
properties, need to be explored deeply, and this motivates us to conduct the present work. Up to
now, the only known multi-rotating charged AdS black hole solutions in seven- and six-dimensional
gauged supergravities are the equal-charge versions constructed by Chow in refs. \cite{CQG25-175010,
CQG27-065004}, while their sisters with two generic unequal electric charges still remain elusive
till now. Ever since Chow \cite{CQG25-175010,CQG27-065004} had constructed the equal-charged
rotating black holes in the seven- and six-dimensional gauged supergravity theories, there are
a lot of attention being paid to various aspects of these solutions. For example, Kerr/CFT
correspondence \cite{PRD79-084018}, hidden symmetry \cite{CQG27-205009}, area product formula
\cite{PRL106-121301}, extremal vanishing horizon limit \cite{JHEP0220154} and gravitational
Cardy limit \cite{JHEP1120041} have been investigated in recent years. In this paper, we will
first construct the ultra-spinning counterpart of the six-dimensional Chow's rotating charged
black hole, then we investigate its thermodynamical properties and show that the obtained
thermodynamical quantities perfectly obey both the first law and the Bekenstein-Smarr mass
formula. Furthermore, we will also investigate wether or not the RII is satisfied in the
singly- and doubly-rotating cases, respectively.

The remaining parts of this article are organized as follows: In sec. \ref{sII}, we first
retrospect the six-dimensional Chow's doubly-rotating charged AdS black hole solution with
two equal electric charges in $\cN = 4$, $SU(2)$ gauged supergravity theory, in particular,
present the Lagrangian of the action, and give the necessary expressions of the metric and
the Abelian gauge potential of the solution. To proceed easy, we will first consider the
simpler singly-rotating case. To do so, we perform the coordinate transformations and get
the singly-rotating Chow's solution, and then take the ultra-spinning limit to obtain the
corresponding ultra-spinning charged black hole solution. Then, we analyze its thermodynamical
properties and demonstrate that the reasonable thermodynamical mass and angular momentum can
also be obtained perfectly by the conformal Ashtekar-Magnon-Das (AMD) method and Abbott-Deser
(AD) method. In the meanwhile, we show that all calculated thermodynamical quantities completely
satisfy the differential first law and the Bekenstein-Smarr mass formula. In addition, we also
show that this ultra-spinning version of the singly-rotating charged Chow's black hole does
not always violate the RII, thus it can be either sub-entropic or super-entropic, depending
upon the ranges of the solution parameters. In sec. \ref{sIII}, we turn to consider the more
general doubly-rotating case. Parallel to the above section, we first obtain the ultra-spinning
counterpart of the doubly-rotating charged Chow's black hole solution and investigate its
thermodynamical properties. We will respectively adopt the AMD and AD methods to calculate
the mass and two angular momenta. We found that both methods can self-consistently give the
same expression for the mass. Furthermore, it has been demonstrated that both the differential
first law and the Bekenstein-Smarr mass formula are completely satisfied. What is more, we will
point out that the six-dimensional doubly-rotating Chow's charged black hole but ultra-spinning
only along one spatial axis may violate or obey the RII, thus it can be either super-entropic
or sub-entropic in the general case. Finally, our paper is concluded in sec. \ref{sIV}.

\section{Ultra-spinning Chow's black hole:
The singly-rotating charged case}\label{sII}

\subsection{Chow's rotating charged black
holes in six-dimensional gauged supergravity}

Almost ten years ago, Chow \cite{CQG27-065004} had obtained a six-dimensional rotating charged
AdS black hole solution with two equal charges in the $\cN = 4$, $SU(2)$ gauged supergravity
theory. Its bosonic fields generally contain a metric graviton field $g_{\mu\nu}$, a two-form
potential $\mathcal{A}$, an Abelian one-form potential $A$ and the gauge potentials of $SU(2)$
Yang-Mills theory as well as one scalar field $\varphi$. After some self-consistent truncations,
its Lagrangian can be expressed by using the exterior differential formalism as follows
\cite{CQG27-065004}:
\bea
\mathcal{L}_6 &=& R\star {1\!\!\!1} -\frac{1}{2}\star d\varphi \wedge d\varphi -X^{-2}\big(\star
 F \wedge F +g^2\star \mathcal{A} \wedge \mathcal{A}\big) -F \wedge F \wedge \mathcal{A} \nn \\
&& -\frac{1}{2}X^4\star \mathcal{F} \wedge \mathcal{F} -\frac{g^2}{3}\mathcal{A} \wedge
 \mathcal{A} \wedge \mathcal{A} +g^2\big(9X^2 +12X^{-2} -X^{-6}\big)\star {1\!\!\!1} \, ,
\eea
where $F = dA$, $\mathcal{F} = d\mathcal{A}$, $X = e^{-\varphi/\sqrt{8}}$
and the gauge coupling constant $g$ is equal to the reciprocal of the cosmological scale $l$.

Omitting the expressions of the scalar field and two-form potential, the metric and the Abelian
gauge potential of the Chow's six-dimensional doubly-rotating charged black hole solution with
two equal charges are given by
\cite{CQG27-065004}
\bea \label{chowbh}
d\bar{s}_6^2 &=& \frac{\sqrt{\bar{H}(r,y,z)}}{\sqrt{(r^2+y^2)(r^2+z^2)}}\Bigg[
 -\frac{(r^2+y^2)(r^2+z^2)\bar{\Delta}(r)}{\bar{H}(r,y,z)^2}\bar{U}^2
 +\frac{(r^2+y^2)(r^2+z^2)}{\bar{\Delta}(r)}dr^2 \nn \\
&& +\frac{(r^2+y^2)(y^2-z^2)}{\bar{F}(y)}dy^2 +\frac{\bar{F}(y)}{(r^2+y^2)(y^2-z^2)}
 \Big(\bar{V} -\frac{qr}{\bar{H}(r,y,z)}\bar{U}\Big)^2 \nn \\
&&  +\frac{(r^2+z^2)(z^2-y^2)}{\bar{K}(z)}dz^2 +\frac{\bar{K}(z)}{(r^2+y^2)(z^2-y^2)}
 \Big(\bar{W} -\frac{qr}{\bar{H}(r,y,z)}\bar{U}\Big)^2\Bigg] \, , \\
\bar{A} &=& \frac{\sqrt{q(2m+q)}\, r}{\bar{H}(r,y,z)}\bar{U} \, , \nn
\eea
where
\bea
&& \bar{\Delta}(r) = \Big(1 +\frac{r^2}{l^2} \Big)(r^2+a^2)(r^2+b^2) -2mr
 +\frac{q}{l^2}\big[2r^3 +(a^2+b^2)r +q\big] \, , \nn \\
&& \bar{F}(y) = -\Big(1-\frac{y^2}{l^2}\Big)(a^2-y^2)(b^2-y^2)\, , \qquad
\bar{K}(z) = -\Big(1-\frac{z^2}{l^2}\Big)(a^2-z^2)(b^2-z^2)\, , \nn \\
&& \bar{H}(r,y,z) = (r^2+y^2)(r^2+z^2) +qr\, , \nn
\eea
in which $m$ and $q$ are the mass and electric charge parameters, respectively, and we denote
in the frame rest at infinity:
\bea
\bar{U} &=& \frac{(l^2-y^2)(l^2-z^2)}{\Xi_1\Xi_2l^4}dt
  -\frac{(a^2-y^2)(a^2-z^2)}{a(a^2-b^2)\Xi_1}d\phi
 -\frac{(b^2-y^2)(b^2-z^2)}{b(b^2-a^2)\Xi_2}d\psi \, , \nn \\
\bar{V} &=& \frac{(r^2+l^2)(l^2-z^2)}{\Xi_1\Xi_2l^4}dt
 -\frac{(r^2+a^2)(a^2-z^2)}{a(a^2-b^2)\Xi_1}d\phi
 -\frac{(r^2+b^2)(b^2-z^2)}{b(b^2-a^2)\Xi_2}d\psi \, , \nn \\
\bar{W} &=& \frac{(r^2+l^2)(l^2-y^2)}{\Xi_1\Xi_2l^4}dt
 -\frac{(r^2+a^2)(a^2-y^2)}{a(a^2-b^2)\Xi_1}d\phi
 -\frac{(r^2+b^2)(b^2-y^2)}{b(b^2-a^2)\Xi_2}d\psi \, , \nn
\eea
with $\Xi_1 = 1-a^2/l^2$ and $\Xi_2 = 1-b^2/l^2$.

In the above solution, one is usually concerned with the under-rotating case ($0 \leq\, a < l$,
$0 \leq\, b < l$) and presumably assume that the solution parameters satisfy $m > 0$ and $q
\geq 0$. The angle coordinates $\phi$ and $\psi$ are canonically normalized with period $2\pi$,
namely, ($0 < \phi < 2\pi$, $0 < \psi < 2\pi$), and the time coordinate $t$ is also canonically
normalized. The other coordinates are: $r$ is the usual radial coordinate ($0 < r < \infty$),
$y$ and $z$ representing the latitudinal coordinates whose ranges are now simply taken as
$-a \leq\, y \leq\, a \leq\, z \leq\, b$, for definiteness. $y$ and $z$ are useful coordinates
for describing the metric on $S^4$, because they parameterize directional cosines in a
rather symmetric manner \cite{CQG23-5323}. In order to see that singularities at $y = z$ and
when $a = b$ can be removed, and that the gauge potential is globally defined, one could replace
$y$ and $z$ with latitudinal spherical polar coordinates (for instance, $y = a\cos\theta_1$ and
$z = b\cos\theta_2$). Below we will obviously demonstrate this in the singly-rotating case.

Notice that the above black hole generally rotates along two different axes, corresponding
to rotation parameters $a$ and $b$. We will first study a simpler singly-rotating case in
which the above black hole has only one rotation parameter $a$. To do so, we first perform
the coordinate transformations: $\phi \to \phi +(a/l^2)t$, $z = b\cos\theta$ with $\theta
\in [0, \pi]$, and then take the $b\rightarrow 0$ limit, we get the singly-rotating solution
in an asymptotic rotating frame at infinity:
\bea \label{sr}
d\tilde{s}_6^2 &=& \frac{\sqrt{\tilde{H}(r,y)}}{\sqrt{r(r^2+y^2)}}\bigg\{
 -\frac{(r^2+y^2)\tilde{\Delta}(r)}{\tilde{H}(r,y)^2}
 \Big(dt -\frac{a^2-y^2}{a\Xi}d\phi\Big)^2 \nn \\
&& +\frac{r^2(r^2+y^2)}{\tilde{\Delta}(r)}dr^2 +\frac{r^2+y^2}{\tilde{K}(y)}dy^2
 +\frac{r^2y^2}{a^2}\big(d\theta^2 +\sin^2\theta\, d\psi^2\big) \nn \\
&& +\frac{\tilde{K}(y)}{r^2+y^2}\bigg[dt -\frac{r^2+a^2}{a\Xi}d\phi
 -\frac{q}{\tilde{H}(r,y)}\Big(dt -\frac{a^2-y^2}{a\Xi}d\phi\Big)\bigg]^2 \bigg\}\, , \\
\tilde{A} &=& \frac{\sqrt{q(2m+q)}}{\tilde{H}(r,y)}
 \Big(dt -\frac{a^2-y^2}{a\Xi}d\phi\Big)\, , \nn
\eea
where
\bea
\tilde{\Delta}(r) &=& \Big(1 +\frac{r^2}{l^2}\Big)(r^2+a^2)r^2 -2mr
 +\frac{q}{l^2}(2r^3 +a^2r +q) \, ,\nn \\
\tilde{K}(y) &=& \Big(1-\frac{y^2}{l^2}\Big)(a^2-y^2) \, , \quad
\tilde{H}(r,y) = (r^2+y^2)r +q \, , \quad
\Xi = 1 -\frac{a^2}{l^2} \, . \nn
\eea
In addition, the two-form potential assumes a very simply expression
\bea
\tilde{\mathcal{A}} = \frac{qy^3}{(r^2+y^2)a^2}\sin\theta\, d\theta \wedge d\psi \, , \nn
\eea
although it has no relation to our discussions and doesn't take any effect on the calculation
of the conserved charges.

One can notice that the angular slice in the above metric (\ref{sr}) has apparently been
factorized out a $S^2$-sphere, however this does not mean that the horizon topology
of the spacetime might be a direct-product $S^2 \times S^2$, rather it still remains a
$S^4$-sphere, which is very clear in the static limit \cite{PRL83-5226} by simply sending
$a = 0$ after letting $y = a\cos\vartheta$.

In the next subsection, we will take the ultra-spinning limit $a \to l$ upon the metric
(\ref{sr}) to obtain its ultra-spinning cousin.

\subsection{The ultra-spinning limit}

Since the metric (\ref{sr}) is already written in an asymptotically rotating frame, we therefore
only need to rename a new azimuthal coordinate $\varphi = \phi/\Xi$ to avoid a singular metric
followed by taking the $a\rightarrow l$ limit. Hence we straightforwardly obtain the following
ultra-spinning black hole solution:
\bea \label{sesr}
d\hat{s}^2 &=& \frac{\sqrt{\hat{H}(r,y)}}{\sqrt{r(r^2+y^2)}}\bigg\{
 -\frac{(r^2+y^2)\hat{\Delta}(r)}{\hat{H}(r,y)^2}
 \Big(dt -\frac{l^2-y^2}{l}d\varphi\Big)^2 +\frac{r^2(r^2+y^2)}{\hat{\Delta}(r)}dr^2 \nn \\
&& +\frac{r^2+y^2}{\hat{K}(y)}dy^2 +\frac{\hat{K}(y)}{r^2+y^2}
 \bigg[dt -\frac{r^2+l^2}{l}d\varphi -\frac{q}{\hat{H}(r,y)}
 \Big(dt -\frac{l^2-y^2}{l}d\varphi\Big)\bigg]^2 \nn \\
&& +\frac{r^2y^2}{l^2}\big(d\theta^2 +\sin^2\theta\, d\psi^2\big) \bigg\} \, ,
\eea
where
\bea
&&\hat{\Delta}(r) = \frac{r^2(r^2+l^2)^2 +2qr^3 +q^2}{l^2} -(2m -q)r \, , \nn \\
&&\hat{K}(y) = \frac{(l^2-y^2)^2}{l^2} \, , \qquad \hat{H}(r,y) = (r^2+y^2)r +q \, . \nn
\eea

Since the new azimuthal coordinate $\varphi$ is noncompact, one can compactify it by requiring:
$\varphi \sim \varphi +\mu$ with a dimensionless positive parameter $\mu$. Also one can easily
find the Abelian gauge potential and the two-form potential in this limit as
\be
\hat{A} = \frac{\sqrt{q(2m+q)}}{\hat{H}(r,y)}\Big(dt -\frac{l^2-y^2}{l}d\varphi\Big)\, , \qquad
\hat{\mathcal{A}} = \frac{qy^3}{(r^2+y^2)l^2}\sin\theta\, d\theta \wedge d\psi\, .
\ee

\subsection{Horizon geometry}

We would like to seek the condition to ensure that the new ultra-spinning solution (\ref{sesr})
indeed describes an AdS black hole, for which one must have
\bea
&& m \ge m_e = \frac{r_e}{2l^2}\big(36r_e^4 +19l^2r_e^2 +3l^4\big) \, , \label{mre} \\
&&\qquad r_e = \bigg(\frac{q +\sqrt{q^2 +4l^6/135}}{10}\bigg)^{1/3}
 -\frac{l^2}{15}\Big(\frac{10}{q +\sqrt{q^2 +4l^6/135}}\Big)^{1/3} \, , \label{res}
\eea
where $r_e > 0 $ is the only real root of the cubic equation: $r_e(5r_e^2 +l^2) = q$,
which is the location of the extremal horizon that satisfies $\hat{\Delta}(r_e) = 0$
and $\hat{\Delta^{\prime}}(r_e) = 0$. For $m > m_e$, there exists an event horizon, while
for $m < m_e$ there is a naked singularity. When $m = m_e$, the black hole becomes extremal.

Now, let us check the geometry of the event horizon, whose location is the largest root
of $\hat{\Delta}(r_+) = 0$. On a constant ($t,r$)-sector, the induced metric on the event
horizon reads
\bea
ds_+^2 &=& \sqrt{\big(r_+^2+y^2\big)\big(r_+^2+y^2 +q/r_+\big)}
 \bigg[\frac{(r_+^2+l^2 +q/r_+)^2(l^2-y^2)^2}{(r_+^2+y^2 +q/r_+)^2l^4}d\varphi^2 \nn \\
&& +\frac{l^2}{(l^2-y^2)^2} dy^2 +\frac{r_+^2y^2}{(r_+^2+y^2)l^2}
 \big(d\theta^2 +\sin^2\theta\, d\psi^2\big)\bigg] \, . \label{hm}
\eea
Apparently the signature of the above metric is strictly Euclidean, so the obtained geometry is
free of any closed timelike curve.

However, the geometry (\ref{hm}) seems to be singular at $y = \pm\,l$. To verify that there is
nothing pathological occurring near these points, let us examine the metric near $y = l$ by
shifting the coordinate: $y = l -\zeta$, then the horizon metric (\ref{hm}) for small $\zeta$
becomes
\be
ds_+^2 \simeq
 \sqrt{\big(r_+^2+l^2\big)\big(r_+^2+l^2+q/r_+\big)}\Big[\frac{d\zeta^2}{4\zeta^2}
 +\frac{4\zeta^2}{l^2}d\varphi^2 +\frac{r_+^2}{r_+^2+l^2}(d\theta^2
 +\sin^2\theta\, d\psi^2)\Big] \, . \label{nhs}
\ee
Clearly, the two-dimensional ($\theta,\psi$)-surface is a standard $S^2$-sphere, while
the ($\zeta,\varphi$)-slice is described by a metric of constant negative curvature on a quotient
of the hyperbolic space $\mathbb{H}^2$, implying that this sector is an $S^2$-sphere with two
punctures. Due to the symmetry, the same is also true for the point at $y = -l$. Thus, there is
no true singularity at these two points. One can easily note that the horizon geometry (\ref{nhs})
reduces to that was studied previously in refs. \cite{PRL115-031101,JHEP0615096} in the uncharged
case when $q = 0$. This implies that our newly obtained black hole solution (\ref{sesr}) also
enjoys a finite area but noncompact horizon.

\subsection{Thermodynamical quantities}\label{sbtq}

Now we turn to investigate some thermodynamical aspects of the above ultra-spinning version
of the six-dimensional singly-rotating Chow's black hole. Although the event horizon of the
above ultra-spinning black hole (\ref{sesr}) is noncompact, it has a finite area and the
Bekenstein-Hawking entropy is one quarter of its horizon area:
\be
\hat{S} = \frac{\hat{A}_+}{4} = \frac{1}{3}\pi\mu\, r_+(r_+^3 +l^2r_+ +q) \, ,
\ee
where $r_+$ is the location of the event horizon, which is the largest root of the equation
$\hat{\Delta}(r_+) = 0$. The Hawking temperature is proportional to the surface gravity $\kappa$
on the event horizon
\be
\hat{T} = \frac{\kappa}{2\pi}
 = \frac{\hat{\Delta}^{\prime}(r_+)}{4\pi\, r_+(r_+^3 +l^2r_+ +q)} \, .
\ee

Note that the above ultra-spinning Chow's black hole is rotating with the speed of light at
infinity, so one must work in a co-rotating frame at the infinity. The angular velocity of
the event horizon is given by
\be
\hat{\Omega} = -\, \frac{g_{t\phi}}{g_{\varphi\varphi}}\Big|_{r = r_+}
 = \frac{r_+l}{r_+^3 +l^2r_+ +q} \, .
\ee

Due to the well falling-off behaviors of the scalar field and the two-form potential as well
as its field strength three-form, the electric charge $Q$ of the black hole can be simply
calculated by using the following Gauss' law integral:
\be
\hat{Q} = \frac{1}{8\pi}\int\star d\hat{A} = \frac{1}{2}\mu\sqrt{q(2m+q)} \, ,
\ee
and its corresponding electrostatic potential at the event horizon reads
\be
\hat{\Phi} = \big(\hat{A}_\mu\chi^\mu \big)\big|_{r = r_+}
 = \frac{\sqrt{q(2m+q)}}{r_+^3 +l^2r_+ +q} \, ,
\ee
where $\chi = \p_t +\hat{\Omega}\p_\varphi$ is the time-like Killing vector normal to the
event horizon, which is also used to define the above surface gravity $\kappa$.

In the following, we will utilize two different methods to calculate the most important
conserved charges: the mass and the angular momentum.

\subsubsection{Conformal Ashtekar-Magnon-Das (AMD) method}

Following the method proposed in ref. \cite{PRD73-104036}, we will first adopt the AMD procedure
\cite{CQG1-L39,CQG17-L17} to calculate the mass and angular momentum of the above asymptotically
AdS spacetime. The AMD mass and angular momentum are evaluated via the integrals in terms of certain
components of the conformal Weyl tensor over the spatial conformal boundary at infinity. The
corresponding asymptotic boundary AdS 5-metric approaches to
\be \label{asesr}
\lim_{r \to \infty}\frac{d\hat{s}^2}{r^2} \approx -\, \frac{1}{l^2}\Big(dt
 -\frac{l^2-y^2}{l}d\varphi\Big)^2+\frac{\hat{K}(y)}{l^2}d\varphi^2 +\frac{dy^2}{\hat{K}(y)}
 +\frac{y^2}{l^2}\big(d\theta^2 +\sin^2\theta\, d\psi^2\big) \, , \quad
\ee
to which one can also define a normal vector: $\hat{n}^a = -l^{-2}r^2 (\p_r)^a$. Notice that
here $\varphi$ is a null coordinate, so the procedure adopted in \cite{PRD73-104036} is much
more suitable for calculating the conserved charges.

Note that the 5-volume form for the conformal boundary AdS metric (\ref{asesr}) is simply given by
\be
\mathbb{V}_5 = \frac{y^2}{l^4}\sin\theta\, dt \wedge dy \wedge
 d\varphi \wedge d\theta \wedge d\psi \, ,
\ee
then by defining $d\Sigma_{\mu} = <\p_{\mu}, \mathbb{V}_5>$ and using the inner-product rule
$<\p_{\mu}, dx^{\mu}> = \delta_{\mu}^{\nu}$, we only need to obtain the $t$-component of the
area vector $d\Sigma_{\mu}$:
\be
d\Sigma_t = \frac{y^2}{l^4}\sin\theta\, dy \wedge d\varphi \wedge d\theta \wedge d\psi \, ,
\ee
which yields the only non-vanishing component: $dS_t = y^2\sin\theta/l^4$. By using the formula:
\be \label{AMDm}
\mathcal{Q}[\xi] = \frac{-l^3}{24\pi}\int_0^{\mu}d\varphi \int_0^{\pi}d\theta \int_0^{2\pi}d\psi
 \int_0^{l}dy \big(r^3C^t_{~cbd}\hat{n}^c\hat{n}^d\xi^b\, dS_t\big)|_{r\to\infty} \, ,
\ee
where $\xi^b$ is a unit Killing vector, we can calculate the conserved mass and the angular
momentum as follows
\be
\hat{M} = \mathcal{Q}[\p_t] = \frac{1}{3}\mu(2m+q) \, , \qquad
\hat{J} = \mathcal{Q}[\p_{\varphi}] = \frac{1}{6}\mu(2m+q)l \label{MJ} \, ,
\ee
among which an explicit ``chirality condition" $\hat{J} = \hat{M}l/2$ follows. Note that the
angular momentum can be also computed by the Komar integral, and the charge parameter $q$ can
be expressed in terms of the conserved charges $\hat{Q}$ and $\hat{M}$ as: $q = 4\hat{Q}^2/
(3\mu\hat{M})$.

\subsubsection{Abbott-Deser (AD) method}

In this subsection, we will alternatively apply the AD method to calculate the mass and the
angular momentum. In the AD procedure \cite{NPB195-76}, the asymptotically AdS spacetime metric
$g_{\mu\nu}$ is separated into a perturbative form:
\be
g_{\mu\nu} = \bar{g}_{\mu\nu} +h_{\mu\nu} \, ,
\ee
where the background metric $\bar{g}_{\mu\nu}$ represents the pure AdS$_6$ spacetime, and
$h_{\mu\nu}$ is the perturbation part. The pure AdS metric is easily obtained by setting
$m = q = 0$ in eq. (\ref{sesr}) as follows:
\bea \label{pAdS6}
d\bar{s}_{\rm AdS}^2 &=& -\, \frac{r^2 -y^2 +2l^2}{l^2}dt^2
 +2\frac{(r^2+l^2)(l^2-y^2)}{l^3}dtd\varphi +\frac{(r^2+y^2)l^2}{(r^2+l^2)^2}dr^2 \nn \\
&& +\frac{r^2+y^2}{\hat{K}(y)}dy^2 +\frac{r^2y^2}{l^2}\big(d\theta^2
 +\sin^2\theta\, d\psi^2\big) \, ,
\eea
with the determinant being $\sqrt{-\bar{g}} = (r^2+y^2)r^2y^2\sin\theta/l^3$.

After performing the background substraction, one can easily get every components of the
perturbation metric tensor $h_{\mu\nu}$, whose leading asymptotic behaviors at the infinity
are not difficult to be obtained via series expansions:
\bea
&& h_{tt} \simeq -\, \frac{q}{2l^2r} +\cO(r^{-2})\, , \quad
 h_{t\varphi} \simeq \frac{q(l^2-y^2)}{2l^3r} +\cO(r^{-2})\, , \quad
 h_{\psi\psi} \simeq \frac{qy^2}{2l^2r}\sin^2\theta +\cO(r^{-2}) \, , \nn \\
&&  h_{xx} \simeq \frac{qy^2}{2l^2r} +\cO(r^{-2}) \, , \qquad
 h_{yy} \simeq \frac{ql^2}{2(l^2-y^2)^2r} +\cO(r^{-4})\, , \nn \\
&& h_{\varphi\varphi} \simeq \frac{(2m+q)(l^2-y^2)^2}{l^2r^3} +\cO(r^{-4}) \, , \qquad
 h_{rr} \simeq -\, \frac{3ql^2}{2r^5} +\cO(r^{-6}) \, ,  \nn
\eea
from which one can be that they are well-behaved at infinity.

Defining two symmetric tensors as did in \cite{NPB195-76}
\be
K^{\mu\nu\rho\sigma} =\bar{g}^{\mu\rho}H^{\nu\sigma} +\bar{g}^{\nu\sigma}H^{\mu\rho}
 -\bar{g}^{\mu\sigma}H^{\nu\rho} -\bar{g}^{\nu\rho}H^{\mu\sigma} \, , \qquad
H^{\mu\nu} = h^{\mu\nu} -\frac{1}{2}\bar{g}^{\mu\nu}h^\rho_{~\rho} \, ,
\ee
and using the formula\footnote{The covariant differential operator $\bar{\nabla}$ is defined
on the background connection $\bar{\Gamma}$ of the pure AdS metric (\ref{pAdS6}).}
\be \label{ADm}
\mathcal{Q}[\xi] = \frac{1}{16\pi}\int_0^{\mu}d\varphi \int_0^{\pi}d\theta
 \int_0^{2\pi}d\psi \int_0^{l}dy \sqrt{-\bar{g}}\big(\xi_c\bar{\nabla}_d K^{trcd}
 +K^{tcdr}\bar{\nabla}_d\xi_c\big)|_{r\to\infty} \, ,
\ee
one can straightforwardly calculate the conserved mass and the angular momentum as follows:
\be
\hat{M} = \mathcal{Q}[\p_t] = \frac{1}{3}\mu(2m+q) \, , \qquad
\hat{J} = \mathcal{Q}[\p_{\varphi}] = \frac{1}{6}\mu(2m+q)l
 = \frac{1}{2}\hat{M}l \, ,
\ee
Obviously, they coincide with the previous results (\ref{MJ}) computed by the conformal AMD method.

It is not difficult to verify that the above thermodynamical quantities completely satisfy both
the differential first law of black hole thermodynamics and Bekenstein-Smarr formula
\bea
 d\hat{M} &=& \hat{T}d\hat{S} +\hat{\Omega}d\hat{J} +\hat{\Phi}d\hat{Q}
 +\hat{\mathcal{V}}dP +\hat{K}d\mu\, , \\
 \hat{M} &=& \frac{4}{3}\big(\hat{T}\hat{S} +\hat{\Omega}\hat{J}\big)
 +\hat{\Phi}\hat{Q} -\frac{2}{3}\hat{\mathcal{V}}P \, ,
\eea
with the thermodynamic volume and a new chemical potential:
\bea
&& \hat{\mathcal{V}} = \frac{1}{5}\Big(4r_+ +\frac{q}{r_+^2}\Big)S_+ \, , \\
&& \hat{K} = -\, (2m+q)\frac{r_+^3 -l^2r_+ +q}{12(r_+^3 +l^2r_+ +q)}
 = -\, \frac{(r_+^3 +q)^2 -l^4r_+^2}{12l^2r_+} \, ,
\eea
which are conjugate to the pressure $P = 5/(4\pi l^2)$ and the dimensionless parameter $\mu$,
respectively. It should be pointed out that all the above thermodynamical quantities can also
be easily obtained by the limit procedure originally suggested in \cite{PRD101-024057}, and
the expressions are also applicable to the inner horizon by replacing $r_+$ to $r_-$. Knowing
this latter fact is very important to our discussions below about RII.

\subsection{RII}

Ten years ago, it was conjectured \cite{PRD84-024037} that the AdS black hole satisfies the
following RII:
\be
\mathcal{R} = \Big[\frac{(D-1)\mathcal{V}}{\omega_{D-2}}\Big]^{1/(D-1)}
 \Big(\frac{\omega_{D-2}}{A}\Big)^{1/(D-2)} \ge 1 \, . \label{ipr}
\ee
Here $\mathcal{V}$ is the thermodynamic volume, $A$ is the horizon area, and $\omega_{D-2}$
stands for the area of the space orthogonal to constant ($t,r$)-surfaces. Equality is attained
for the Schwarzschild-AdS black hole, which implies that the Schwarzschild-AdS black hole has
the maximum entropy. In other words, for a given entropy, the Schwarzschild-AdS black hole owns
the least volume, thus it is most efficient in storing information.

It is worthy to examine whether the above ultra-spinning solution of the six-dimensional
singly-rotating Chow's black hole violates the RII or not. Note that, due to the compatification
of $\varphi$-axis, the volume of the $S^4$-sphere in this $D = 6$ spacetime is: $\omega_4 =
4\pi\mu/3$. Substituting the thermodynamic volume: $\mathcal{V} = \hat{\mathcal{V}} = (4r_+^3
+q)A_+/(20r_+^2)$ and the horizon area: $A = \hat{A}_+ = 4\pi\mu\, r_+(r_+^3 +l^2r_+ +q)/3$
into the above isoperimetric ratio (\ref{ipr}), we now get
\be
\mathcal{R} = \frac{\big(1 +qr_+^{-3}/4\big)^{1/5}}{
 \big(1 +l^2r_+^{-2} +qr_+^{-3}\big)^{1/20}} \, . \label{irs}
\ee

In the uncharged AdS black hole case ($q = 0$), the above ratio always complies with the
inequalities: $0 \leq \mathcal{R} < 1$, and the black hole is always super-entropic
\cite{PRL115-031101}. In the charged case ($q > 0$), the conclusion will change drastically,
and we will show that this ratio can be equal to, larger than or small than unity, so the new
ultra-spinning charged black hole can be either sub-entropic or super-entropic, depending upon
the ranges of the mass and charge parameters. This property is remarkably different from that
of the singly-rotating Kerr-AdS$_6$ super-entropic black hole \cite{PRL115-031101,JHEP0615096},
which always strictly violates the RII. By the way, the ratio $\mathcal{R}$ is monotonically
decreasing with $r_+/l$ when $q > 0$. Below, we will provide some numerical evidences to support
these conclusions.

Before performing numerical analysis, we first need to address the ranges of the solution
parameters ($m, q$) that characterize our new solution (\ref{sesr}). Without loss of generality,
we will set $l = 1$ and assume that $\mu > 0$ from now on, when presenting numerical results.
By virtue of the positiveness of the conserved mass $\hat{M}$, it demands that the condition:
$q > -2m$ must be satisfied; then the rationality of the electric charge $\hat{Q}$ further
requires that $q \geq 0$. Otherwise, if $q = 4\hat{Q}^2/(3\mu\hat{M}) < 0$, then one must
confront with the situation where the mass $\hat{M}$ is negative. Therefore, it is suffice
to choose both the mass and charge parameters to be positive: $q \geq 0$ and $m > 0$.

Keeping in mind that although we are discussing about the outer horizon, the same is also
applicable to the inner horizon. So only portions of curves correspond to the outer horizon.
Then it is an important matter to mark out the locations of extremal horizons on the plotted
curves. Let us begin with by determining the horizon radii, which are the real roots of the
horizon equation: $\hat{\Delta}(r_h) = 0$, a sextic polynomial equation of $r_h$. Instead by
solving $r_h$, rather we can express the mass parameter $m$ as a function of $r_h$ and $q$:
\bea
m = \frac{r_h^2(r_h^2+l^2)^2 +qr_h(2r_h^2+l^2) +q^2}{2l^2r_h} \, . \nn
\eea

\begin{figure}[!h]
\centering
\subfigure[\, $m$ vs $r_h$ for $q = (0, 0.25, 0.5, 0.75)$]{\label{mrq}
\includegraphics[width=0.45\textwidth,height=0.25\textheight]{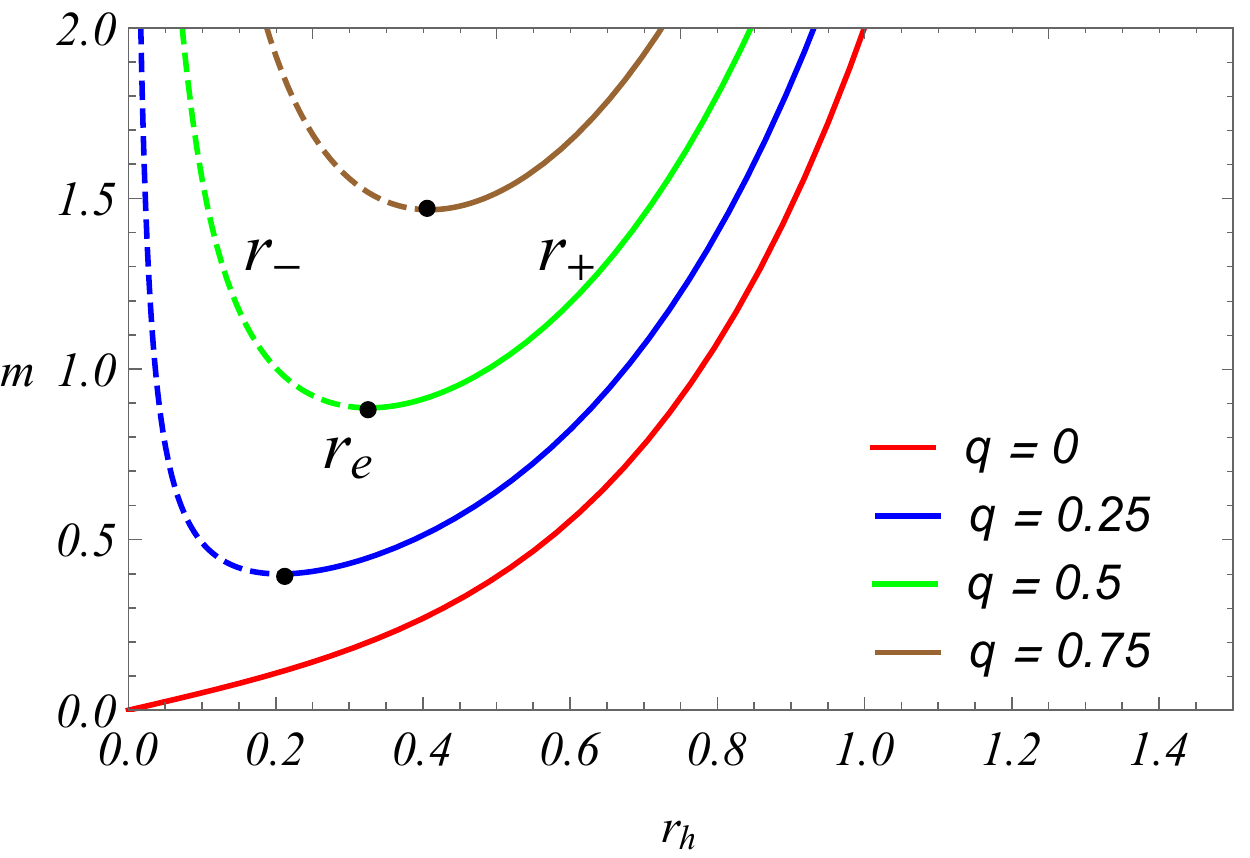}}
\hspace{1cm}
\subfigure[\, $\mathcal{R}$ vs $r_h$ for $q = (0, 0.25, 0.5, 0.75)$]{\label{Rrq}
\includegraphics[width=0.45\textwidth,height=0.25\textheight]{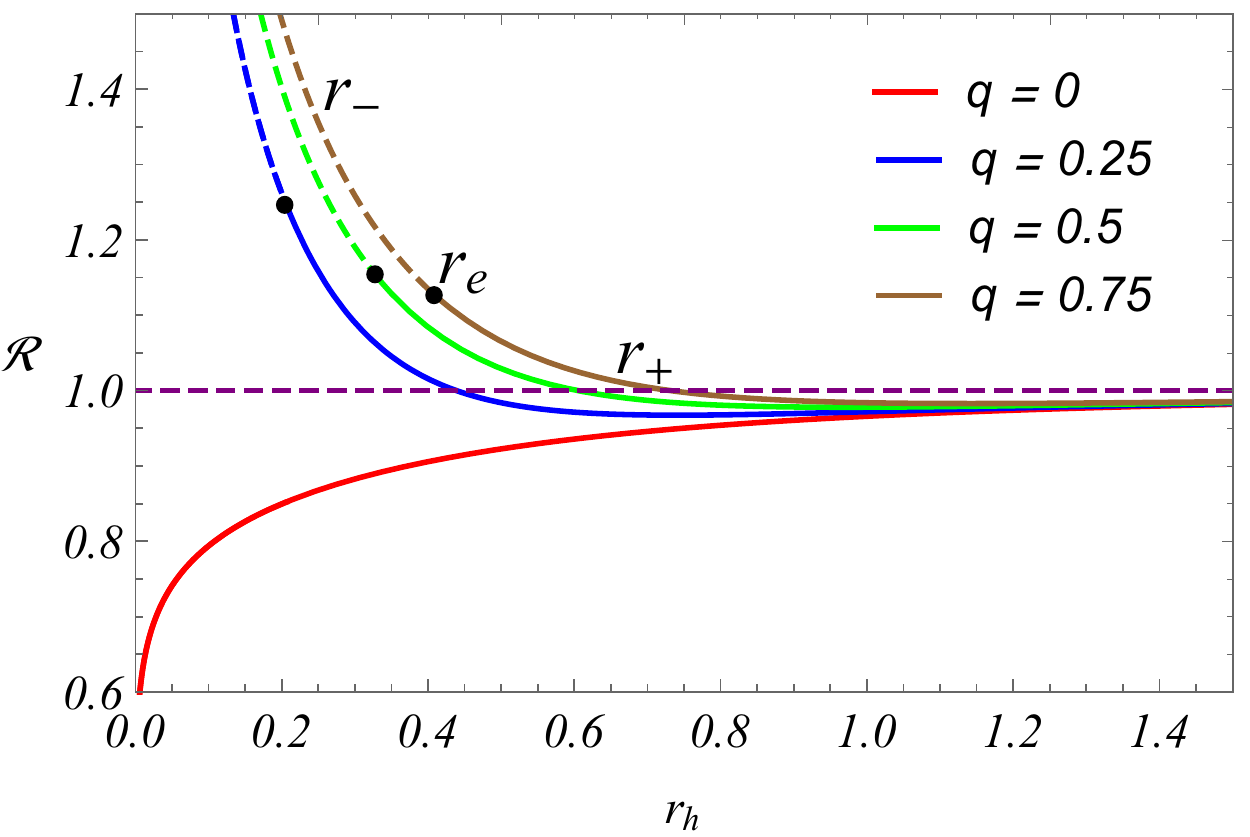}}
\caption{The mass parameter $m$ and isoperimetric ratio $\mathcal{R}$ vs horizon radius $r_h$
and different numerical values of the electric charge parameter $q$ with $l = 1$. Solid/dashed
lines correspond to outer/inner horizons, while black dots mark out the extremal horizons.
Dashed horizontal purple line corresponds to $\mathcal{R} = 1$.}
\label{Rs1}
\end{figure}

In Fig. \ref{Rs1}, we plot the mass $m$ and the isoperimetric ratio $\mathcal{R}$ as functions
of the horizon radius by setting the charge parameter to some fixed numerical values: $q = (0,
0.25, 0.5, 0.75)$. When $q = 0 = r_e(5r_e^2 +l^2)$, it is clear that there is no extremal horizon,
as is also obviously shown in Fig. \ref{mrq}, and both red-marked curves in Fig. \ref{Rs1} start
from the coordinate origin. But when $q\neq 0$, there is a possibility that the horizon is degenerate
in the extremal black hole case, and is given by eq. (\ref{res}). A very key point is to specify
where are the concrete locations of these extremal horizons on the plotted curves. In the extremal
case for $q = (0.25, 0.5, 0.75)$, the corresponding numerical values for the extremal horizon
radius $r_e$, mass parameter $m_e$ and isoperimetric ratio $\mathcal{R}_e = \mathcal{R}(r_e)$
are computed via eqs. (\ref{res}), (\ref{mre}) and (\ref{irs}), respectively, and tabulated in
Tab. \ref{exT1} and marked out by the black dots in Fig. \ref{Rs1}.

\begin{table*}[h]
\centering
\begin{tabular}{c|c|c|c}\hline
$q  $ & $r_e$       & $m_e$       & $\mathcal{R}_e$ \\ \hline
0.25  & 0.206177706 & 0.399235187 & 1.246797394 \\ \hline
0.5   & 0.326296703 & 0.886060115 & 1.155596682 \\ \hline
0.75  & 0.408689322 & 1.466753380 & 1.127194601 \\ \hline
\end{tabular}
\caption{The numerical values of the horizon radius $r_e$, mass parameter $m_e$, and
isoperimetric ratio $\mathcal{R}_e$ with different electric charge parameter $q$ in
the extremal black hole case ($l = 1$). \label{exT1}}
\end{table*}

Now it is ready for us to explain what these two graphs can tell about. For a given charge
parameter $q > 0$ (say, for example, $q = 0.25$), each curve in Fig. \ref{mrq} has a minimum
$m_e$ at the extremal horizon radius $r_e$, which is the only, double root of $\hat{\Delta}(r_e)
= 0$ and $\hat{\Delta^{\prime}}(r_e) = 0$. Because of the mass parameter $m > m_e$, then
increasing $m$ means that there will be two distinct roots $r_h = r_{\pm}$ of horizon equation:
$\hat{\Delta}(r_h) = 0$, and the larger/smaller root corresponds to the outer/inner horizon,
respectively. Accordingly, the solid curve correspond to the regime $r_+ > r_e$, while the
dashed one to $r_- < r_e$. So each curve is segmented by $r_e$ into two pieces, and only the
$r_+ > r_e$ piece of each curve is really in one-to-one correspondence to the event horizon.
These explanations completely apply to Fig. \ref{Rrq}, where a dashed horizontal purple line
represents the ratio $\mathcal{R} = 1$. The reason is very simple: for a given $q > 0$, there
must exist one and only one point of each curve that exactly corresponds to $r_e$.

It is easy to see from Fig. \ref{mrq} that the mass parameter $m$ is always positive. This is
not only in accordance with our initial assumption, but also reflects the fact that the positive
mass theorem is totally abided. From Fig. \ref{Rrq}, it can be seen that for $q = 0$, one always
has: $0 \leq \mathcal{R} < 1$; but when $q > 0$, three cases: $\mathcal{R} > 1$, $\mathcal{R} =
1$ and $\mathcal{R} < 1$ are all possible. In other words, the singly-rotating Kerr-AdS$_6$
super-entropic black hole always strictly violates the RII, as depicted by a red curve in Fig.
\ref{Rrq} for the uncharged case ($q = 0$). By contrast, in the charged case when $q > 0$, the
isoperimetric ratio $\mathcal{R}$ can be larger/smaller than unity in the regime of smaller/larger
enough $r_+$ (i.e., both $m$ and $q$ are smaller/larger enough). Thus, the six-dimensional
ultra-spinning Chow's black hole with only one rotation parameter can obey or violate the RII,
and it is either sub-entropic or super-entropic, depending upon the ranges of the solution
parameters $m$ and $q$. This character is very similar to that of the ultra-spinning Kerr-Sen-AdS$_4$ black hole previously studied in refs. \cite{PRD102-044007,PRD103-044014}.

\section{Ultra-spinning Chow's black hole:
The doubly-rotating charged case}\label{sIII}

After a warmup excise in the simpler singly-rotating case, we will consider the more general
doubly-rotating case, namely, the six-dimensional ultra-spinning Chow's black hole contain
two different rotation parameters $a$ and $b$.

\subsection{The ultra-spinning limit}

Supposed that we would like to boost the doubly-rotating charged AdS black hole to the speed
of light at the $\psi$-axis, that is, the ultra-spinning direction is now along the $\psi$-axis,
we first need to make a coordinate transformation: $\psi \to \psi +(b/l^2)t$ in the solution
(\ref{chowbh}), so that the metric is written in an asymptotic mixed frame that is rotating
along the $\psi$-axis but rest relative to the $\phi$-axis at infinity. Then we should rename
the azimuthal coordinate $\eta = \psi/\Xi_2$ to avoid a singular metric before taking the
ultra-spinning $b\rightarrow l$ limit while keeping $a$ unchanged. Now by simply setting
$b = l$, we can straightforwardly obtain the following doubly-rotating charged AdS black hole
solution that is ultra-spinning along the $\eta$-axis:
\bea \label{semr}
ds^2 &=& \frac{\sqrt{H(r,y,z)}}{\sqrt{(r^2+y^2)(r^2+z^2)}}\Bigg[
 -\frac{(r^2+y^2)(r^2+z^2)\Delta(r)}{H(r,y,z)^2}U^2
 +\frac{(r^2+y^2)(r^2+z^2)}{\Delta(r)} dr^2 \nn \\
&& +\frac{(r^2+y^2)(y^2-z^2)}{F(y)}dy^2 +\frac{F(y)}{(r^2+y^2)(y^2-z^2)}
 \Big(V -\frac{qr}{H(r,y,z)}U\Big)^2 \nn \\
&& +\frac{(r^2+z^2)(z^2-y^2)}{K(z)}dz^2 +\frac{K(z)}{(r^2+y^2)(z^2-y^2)}
 \Big(W -\frac{qr}{H(r,y,z)}U\Big)^2\Bigg]\, , \\
A &=& \frac{\sqrt{q(2m+q)}\, r}{H(r,y,z)}U \, , \nn
\eea
where $H(r,y,z) = (r^2+y^2)(r^2+z^2) +qr$ still remains unchanged, but now
\bea
&& \Delta(r) = \frac{(r^2+a^2)(r^2+l^2)^2 +q(2r^2
 +a^2 +l^2)r +q^2}{l^2} -2mr \, ,\nn \\
&& F(y) = -\, \frac{(l^2-y^2)^2(a^2-y^2)}{l^2}\, , \quad
 K(z) = -\, \frac{(l^2-z^2)^2(a^2-z^2)}{l^2}\, , \quad
 \Xi = 1 -\frac{a^2}{l^2} \, , \nn
\eea
and
\bea
&&U = dt -\frac{(a^2-y^2)(a^2-z^2)}{a(a^2-l^2)\Xi}\Big(d\phi -\frac{a}{l^2}dt\Big)
 -\frac{(l^2-y^2)(l^2-z^2)}{l(l^2-a^2)}d\eta \, ,\nn \\
&&V = dt -\frac{(r^2+a^2)(a^2-z^2)}{a(a^2-l^2)\Xi}\Big(d\phi -\frac{a}{l^2}dt\Big)
 -\frac{(r^2+l^2)(l^2-z^2)}{l(l^2-a^2)}d\eta \, , \nn \\
&&W = dt -\frac{(r^2+a^2)(a^2-y^2)}{a(a^2-l^2)\Xi}\Big(d\phi -\frac{a}{l^2}dt\Big)
 -\frac{(r^2+l^2)(l^2-y^2)}{l(l^2-a^2)}d\eta \, . \nn
\eea
We now compactify the new azimuthal coordinate $\eta$ via: $\eta \sim \eta +\mu$ with a
dimensionless positive parameter $\mu$, due to its non-compactness. Since the ranges of
the latitudinal coordinates $y$ and $z$ are now taken as $-a \leq\, y \leq\, a \leq\, z
\leq\, l$, so one should examine whether the geometry of the event horizon at constant
($t, r$)-slices is singular near $z \simeq\pm l$. However, it is more complicated to fix
the horizon topology than the singly-rotating case, so we will not discuss this aspect
anymore.

\subsection{Thermodynamical quantities}

As did in the subsection \ref{sbtq}, one can use the same means to evaluate all the thermodynamical
quantities for the above-obtained solution (\ref{semr}). Without repeating any detail of computations,
we simply list the final expressions for two angular velocities, Bekenstein-Hawking entropy, Hawking
temperature, electrostatic potential of the event horizon, and the electric charge as follows:
\bea
&& \Omega_1 = \frac{a\Xi}{B_+}(r_+^2+l^2) +\frac{a}{l^2}
 = \frac{a}{B_+ l^2}\big[(r_+^2+l^2)^2 +qr_+\big] \, ,
\qquad \Omega_2 = \frac{l}{B_+}(r_+^2+a^2) \, , \\
&&S = \frac{A_+}{4} = \frac{\pi\mu}{3\Xi}B_+ \, , \quad
 T = \frac{\Delta^{\prime}(r_+)}{4\pi\, B_+} \, , \quad
 \Phi = \sqrt{q(2m+q)}\frac{r_+}{B_+} \, , \quad
 Q = \frac{\mu}{2\Xi}\sqrt{q(2m+q)} \, ,
\eea
where $B_+ = (r_+^2+a^2)(r_+^2+l^2) +qr_+$.

Similarly, we will also apply two different methods to calculate the mass and two angular momenta.

\subsubsection{
Conformal AMD mass and angular momenta}

In order to apply the conformal AMD method to calculate the AMD mass and angular momenta, first
of all, one needs to get the correct conformal boundary metric for the new ultra-spinning version
(\ref{semr}) of the doubly-rotating charged AdS$_6$ black hole. This boundary AdS metric is
chosen to make its $\phi$-axis to be in a rest frame at infinity, so that one could obtain the
correct expressions for the conserved charges. Just as the same one for the ultra-spinning
uncharged case, the asymptotically AdS boundary metric can be readily obtained by
\bea
\lim_{r \to \infty}\frac{ds^2}{r^2} &=&
 \bigg[-1 +\frac{(a^2-y^2)(a^2-z^2)}{l^4\Xi^2}\bigg]\frac{dt^2}{l^2}
 -\frac{(a^2-y^2)(a^2-z^2)}{a^2l^2\Xi^2}d\phi^2 \nn \\
&& +2\frac{(l^2-y^2)(l^2-z^2)}{l^5\Xi}dtd\eta +\frac{y^2-z^2}{F(y)}dy^2
 +\frac{z^2-y^2}{K(z)}dz^2\, ,
\eea
with which the normal vector: $\hat{n}^a = -l^{-2}r^2 (\p_r)^a$ can be straightforwardly defined.

Notice that the 5-volume element for the above boundary metric is written as
\be
\mathbb{V}_5 = \frac{y^2-z^2}{al^4\Xi^2}dt \wedge dy \wedge dz \wedge d\phi \wedge d\eta \, ,
\ee
so the necessary $t$-component of the area vector $d\Sigma_{\mu}$ can be easily computed:
\be
d\Sigma_t = \frac{y^2-z^2}{al^4\Xi^2}dy \wedge dz \wedge d\phi \wedge d\eta \, .
\ee

As before, by using the formulae (\ref{AMDm}), the conserved AMD mass and two angular momenta is
yielded as follows:
\bea
&&M = \mathcal{Q}[\p_t] = \frac{\mu(1+\Xi)(2m+q\Xi)}{6\Xi^2} \, , \\
&&J_1 = \mathcal{Q}[\p_\phi] = \frac{\mu\, ma}{3\Xi^2} \, , \qquad
 J_2 = \mathcal{Q}[\p_\eta] = \frac{\mu(2m+q\Xi)l}{6\Xi} = \frac{Ml\Xi}{1+\Xi} \, .
\eea
One can note that the previous ``chirality condition": $\hat{J} = \hat{M}l/2$ in the singly-rotating
case is now no longer true in the ultra-spinning $\eta$-axis of the present doubly-rotating
case. In other words, adding extra rotation would break the ``chirality condition".

\subsubsection{AD mass and angular momenta}

Next, we also utilize the AD method to calculate the AD mass and two angular momenta. The
pure AdS metric is obtained by setting $m = q =0$ in the metric (\ref{semr}) as follows:
\bea
d\hat{s}_{AdS}^2 &=& -\, \bigg\{\Big[1 -\frac{(a^2-y^2)(a^2-z^2)}{l^4\Xi^2}
 \Big]r^2 +\frac{l^4}{a^2} -\frac{a^2}{l^4}\Big(\frac{l^4}{a^2}-\frac{l^2-y^2}{\Xi}\Big)
 \Big(\frac{l^4}{a^2}-\frac{l^2-z^2}{\Xi}\Big)\bigg\}\frac{dt^2}{l^2} \nn \\
&& +2\frac{(r^2+l^2)(l^2-y^2)(l^2-z^2)}{l^5\Xi}dtd\eta
 -\frac{(r^2+a^2)(a^2-y^2)(a^2-z^2)}{a^2l^2\Xi^2}d\phi^2 \nn \\
&&+\frac{(r^2+y^2)(r^2+z^2)l^2}{(r^2+a^2)(r^2+l^2)^2}dr^2
 +\frac{(r^2+y^2)(y^2-z^2)}{F(y)}dy^2 +\frac{(r^2+z^2)(z^2-y^2)}{K(z)}dz^2 \, ,
\eea
After background substraction, the leading asymptotic behaviors of the coordinate components of
the perturbative metric tensor $h_{\mu\nu}$ at infinity are given by power series expansion at
larger $r$:
\bea
&&h_{tt} \simeq \frac{q[(a^2-y^2)(a^2-z^2) -l^4\Xi^2]}{2l^6\Xi^2r} +\cO(r^{-3}) \, , \qquad
 h_{t\eta} \simeq \frac{q}{2l^5\Xi\, r}(l^2-y^2)(l^2-z^2) +\cO(r^{-3}) \, , \nn \\
&&h_{t\phi} \simeq \frac{(a^2-y^2)(a^2-z^2)}{al^6\Xi^4r^3}\big[q\Xi(l^2-y^2)(l^2-z^2)
 +2ml^4\Xi^2 -2m(a^2-y^2)(a^2-z^2)\big] +\cO(r^{-5}) \, , \nn \\
&&h_{yy} \simeq \frac{ql^2(y^2-z^2)}{2(l^2-y^2)^2(y^2-a^2)r} +\cO(r^{-3}) \, , \qquad
 h_{zz} \simeq \frac{ql^2(z^2-y^2)}{2(l^2-z^2)^2(z^2-a^2)r} +\cO(r^{-3}) \, , \nn \\
&&h_{rr} \simeq -\, \frac{3ql^2}{2r^5} +\cO(r^{-7}) \, , \qquad
 h_{\phi\phi} \simeq -\, \frac{q}{2a^2l^2\Xi^2r}(a^2-y^2)(a^2-z^2) +\cO(r^{-3}) \, , \nn \\
&&h_{\phi\eta} \simeq -\,\frac{2m}{al^5\Xi^3\, r^3}
 (a^2-y^2)(l^2-y^2)(a^2-z^2)(l^2-z^2) +\cO(r^{-5}) \, , \nn \\
&&h_{\eta\eta} \simeq \frac{(2m+q)l^2 -qa^2}{l^8\Xi^2r^3}(l^2-y^2)^2(l^2-z^2)^2
 +\cO(r^{-5}) \, . \nn
\eea
It can be seen that they are well-behaved at infinity.

Using the formulae (\ref{ADm}), one can straightforwardly compute the conserved AD mass and
two angular momenta as follows:
\bea
&&M = \mathcal{Q}[\p_t] = \frac{\mu(1+\Xi)(2m+q\Xi)}{6\Xi^2} \, , \quad \\
&& J_1 = \mathcal{Q}[\p_\phi] = \frac{\mu\, ma}{3\Xi^2} \, , \quad
 J_2 = \mathcal{Q}[\p_\eta] = \frac{\mu(2m+q\Xi)l}{6\Xi} \, , \qquad
\eea
which are obviously accordant with the results obtained in the last subsection.

It is not difficult to certify that the above thermodynamical quantities completely obey both
the differential and the integral mass formulae:
\bea
&&dM = TdS +\Omega_1dJ_1 +\Omega_2dJ_2 +\Phi\, dQ +\mathcal{V}dP +Kd\mu \, , \\
&&M  = \frac{4}{3}(TS +\Omega_1J_1 +\Omega_2J_2) +\Phi\, Q -\frac{2}{3}\mathcal{V}P \, ,
\eea
by introducing the thermodynamic volume
\bea
\mathcal{V} &=& \frac{2\pi\mu\, ma^2}{15\Xi^2} +\frac{\pi\mu}{15\Xi}
 \Big(4r_+ +\frac{q}{r_+^2} -\frac{qa^2l^2}{r_+^2}B_+ -\frac{q^2a^2l^2}{r_+B_+^2}\Big)B_+ \nn \\
 &=& \frac{\pi\mu}{15\Xi^2r_+}\Big(\big[(1+3\Xi)r_+^2 +a^2\big]B_+
  +qr_+(r_+^2+l^2) +q^2 -\frac{q^2a^2l^2\Xi}{B_+}\Big) \, , \label{tv}
\eea
and a new chemical potential
\bea
K = -\frac{2m+q\Xi}{12\Xi\, B_+}\big[(r_+^2+a^2)(r_+^2-l^2) +qr_+\big]
 = \frac{q^2a^2}{6\Xi\, r_+B_+} -\frac{(r_+^2+a^2)(r_+^4 -l^4) +2qr_+^3
 +q^2}{12l^2\Xi\, r_+} \, , \nn
\eea
which are conjugate to the pressure $P = 5/(4\pi\, l^2)$ and the variable $\mu$, respectively.

\subsection{RII}

Finally, we would like to check whether the ultra-spinning version of the six-dimensional
doubly-rotating Chow's charged black hole violate the RII or not. After substituting the
second expression for the thermodynamic volume (\ref{tv}), the horizon area $A_+ = 4\pi\mu\,
B_+/(3\Xi)$, and $\omega_4 = 4\pi\mu/3$ into the isoperimetric ratio (\ref{ipr}), now it reads
\bea
\mathcal{R} &=& \frac{\omega_4^{1/20}(5\mathcal{V})^{1/5}}{A_+^{1/4}}
 = \Big(\frac{15\mathcal{V}}{4\pi\mu}\Big)^{1/5}\Big(\frac{\Xi}{B_+}\Big)^{1/4} \nn \\
 &=& \bigg(\frac{\big[(1+3\Xi)r_+^2 +a^2\big]B_+ +qr_+(r_+^2+l^2) +q^2
  -q^2a^2l^2\Xi/B_+}{4\Xi^2r_+}\bigg)^{1/5}\Big(\frac{\Xi}{B_+}\Big)^{1/4} \, , \label{ipro}
\eea
which approaches to a finite limit: $[(1 +3\Xi)/4]^{1/5}/\Xi^{3/20} \geq 1$ at $r_+ = \infty$,
provided that $0 < \Xi \leq 1$.

Clearly, due to the complicated expression of the thermodynamic volume $\mathcal{V}$ since
now the solution is also characterized by another rotation parameter $a$, it is very hard to
analytically judge when the value of $\mathcal{R}$ is equal to, larger than or less than unity.
Therefore, some numerical results are enough to illuminate the issue.

Let us first specify the ranges of the solution parameters ($a, m, q$). Consider only the
under-rotating case ($0 \leq\, a < l$) so that $0 < \Xi \leq 1$, then the positive mass theorem
claims that $q > -2m/\Xi$ (supposed also that $\mu > 0$). On the other hand, the rationality
of the electric charge $Q$ asserts that either (i): both inequalities $q \geq 0$ and $q > -2m$
are simultaneously satisfied; or (ii): both $q < 0$ and $q < -2m$ are obeyed in the meanwhile.
In the case (i), it requires that $q \geq 0$ and $m > 0$. In the case (ii), there is only a
very narrow interval: $-2m/\Xi < q < -2m$, so that the charge parameter is negative: $q < 0$
while $m > 0$. To be in conformity with the singly-rotating case discussed in the last section,
below we will only copy with the realistic case where both $m > 0$ and $q \geq 0$, and simply
exclude the odd case (ii).

\begin{figure}[!h]
\centering
\subfigure[\, $\mathcal{R}$ vs $r_h$ for $a = 0.3$, $q = (0, 0.25, 0.5, 0.75)$]{\label{Rrqa3}
\includegraphics[width=0.45\textwidth,height=0.25\textheight]{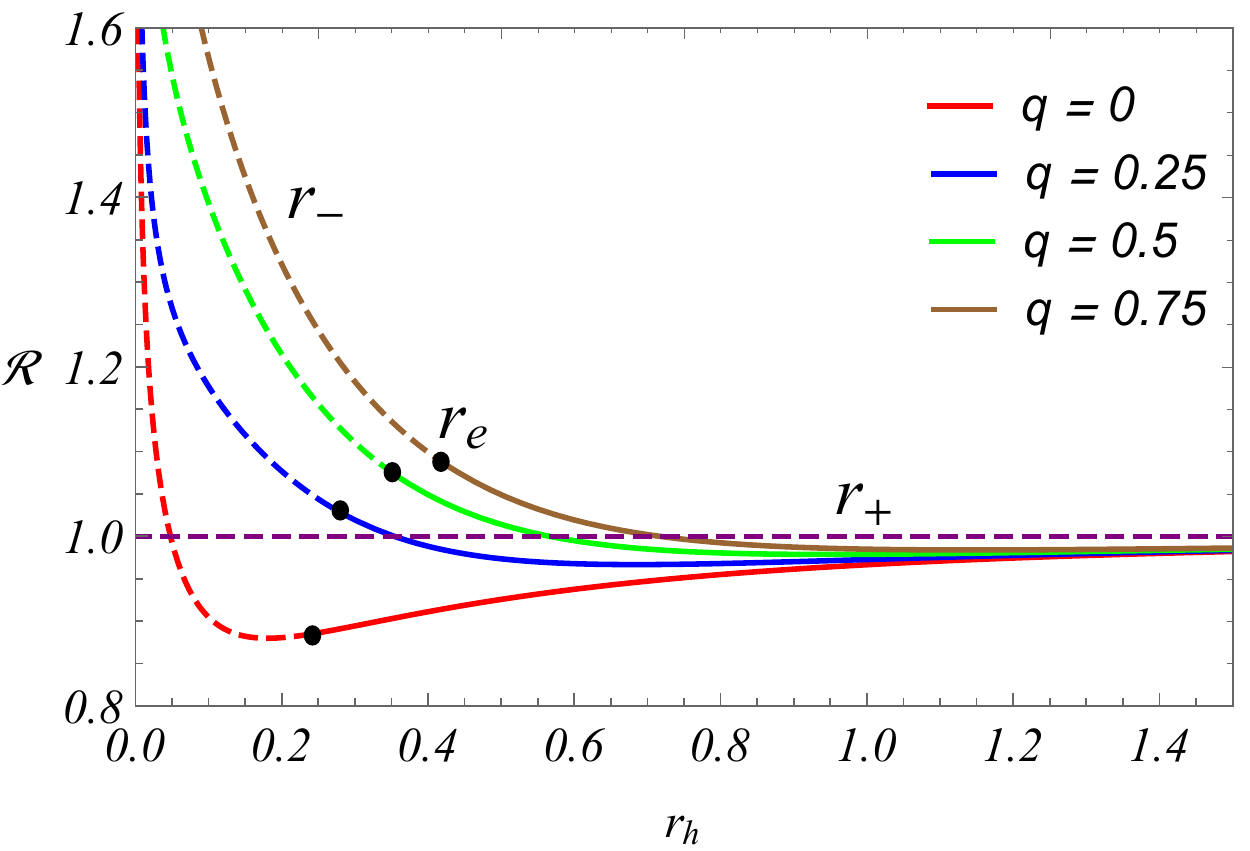}}
\hspace{1cm}
\subfigure[\, $\mathcal{R}$ vs $r_h$ for $a = 0.6$, $q = (0, 0.25, 0.5, 0.75)$]{\label{Rrqa6}
\includegraphics[width=0.45\textwidth,height=0.25\textheight]{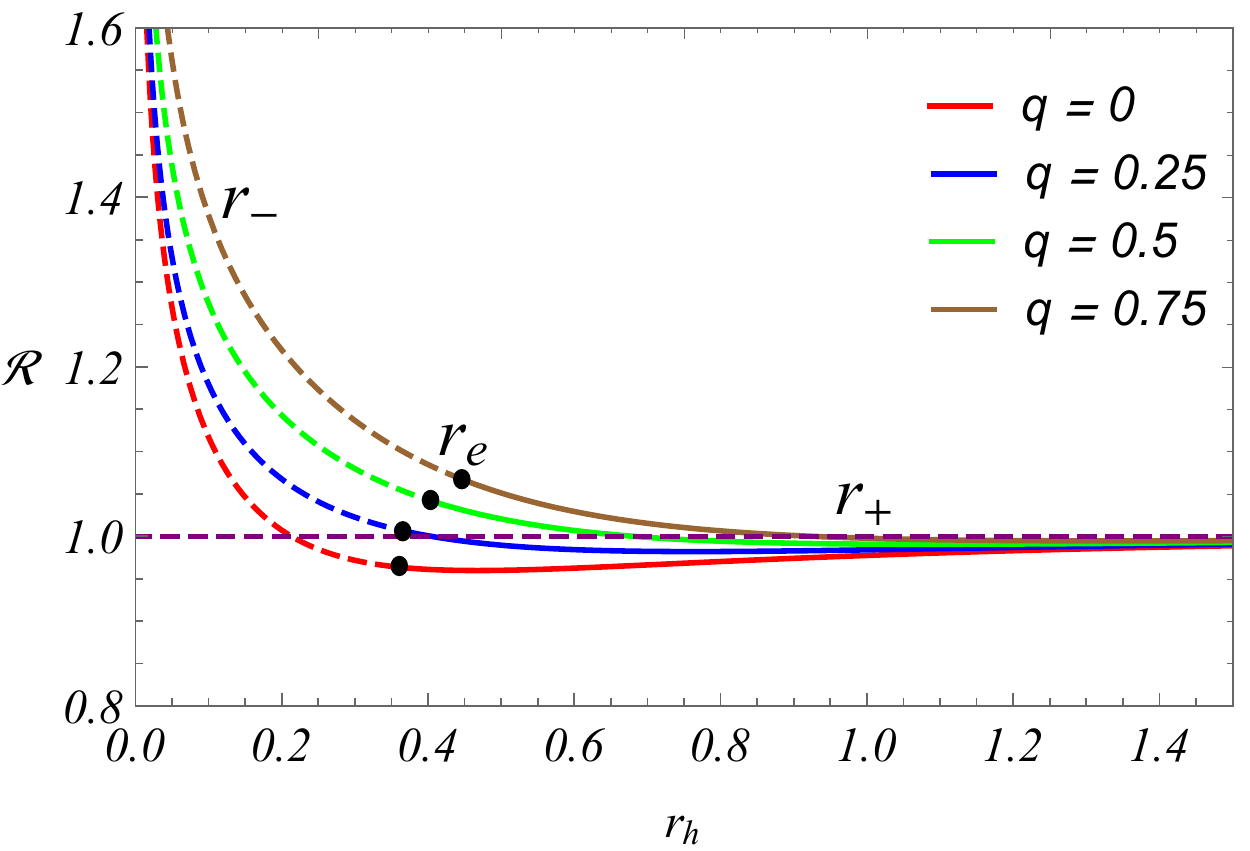}}
\end{figure}

\begin{figure}[!h]
\centering
\subfigure[\, $\mathcal{R}$ vs $r_h$ for $a = 0.9$, $q = (0, 0.25, 0.5, 0.75)$]{\label{Rrqa9}
\includegraphics[width=0.45\textwidth,height=0.25\textheight]{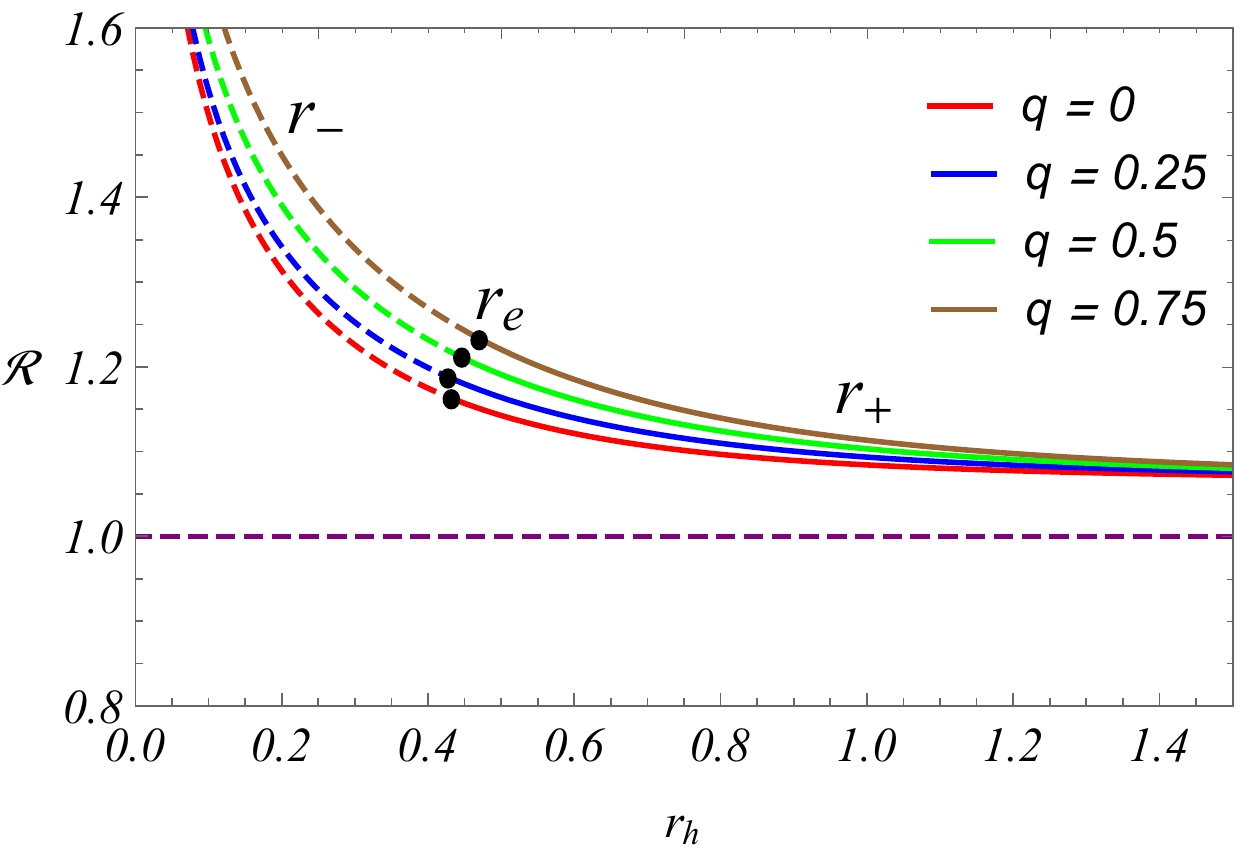}}
\hspace{1cm}
\subfigure[\, $\mathcal{R}$ vs $r_h$ for $q = 0.25$, $a = (0.3, 0.6, 0.9)$]{\label{Rraq25}
\includegraphics[width=0.45\textwidth,height=0.25\textheight]{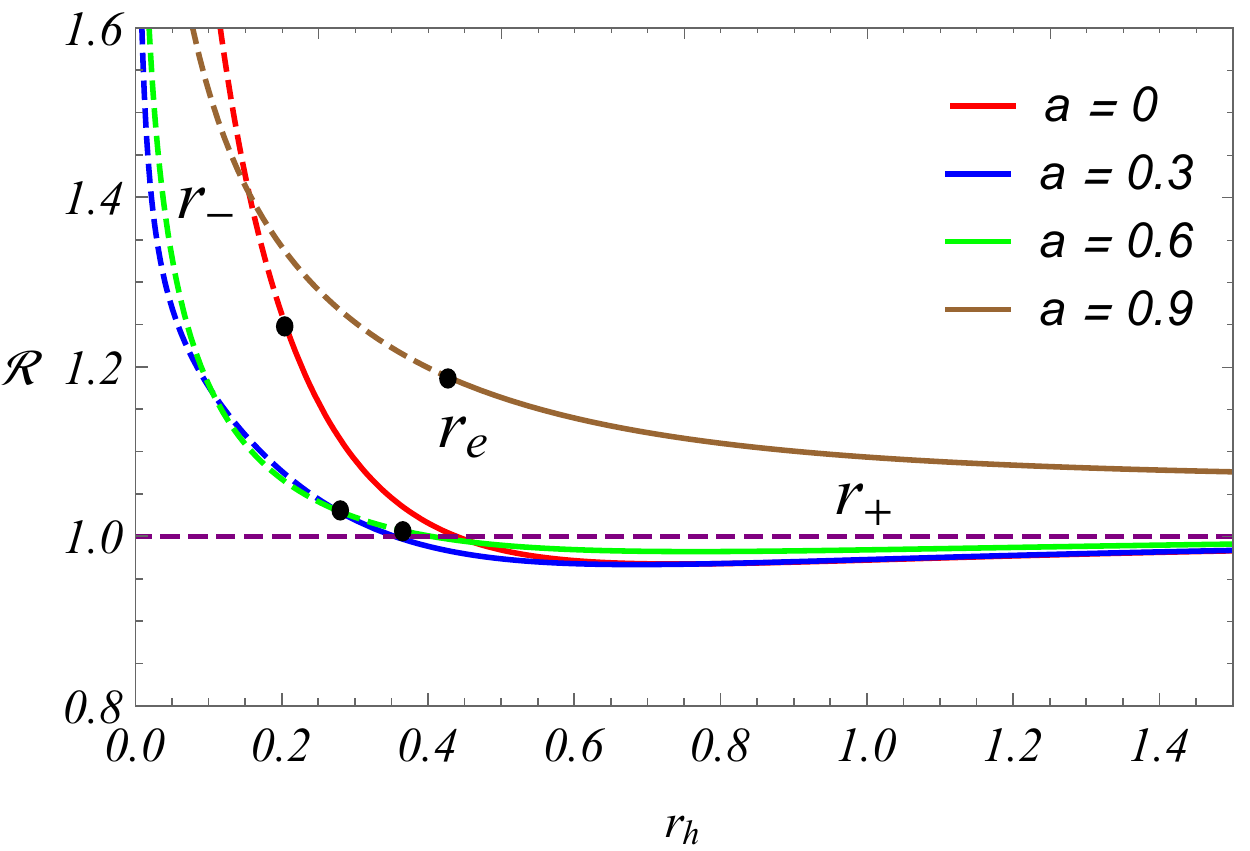}}
\end{figure}

\begin{figure}[!h]
\centering
\subfigure[\, $\mathcal{R}$ vs $r_h$ for $q = 0.5$, $a = (0.3, 0.6, 0.9)$]{\label{Rraq50}
\includegraphics[width=0.45\textwidth,height=0.25\textheight]{Rraq025.pdf}}
\hspace{1cm}
\subfigure[\, $\mathcal{R}$ vs $r_h$ for $q = 0.75$, $a = (0.3, 0.6, 0.9)$]{\label{Rraq75}
\includegraphics[width=0.45\textwidth,height=0.25\textheight]{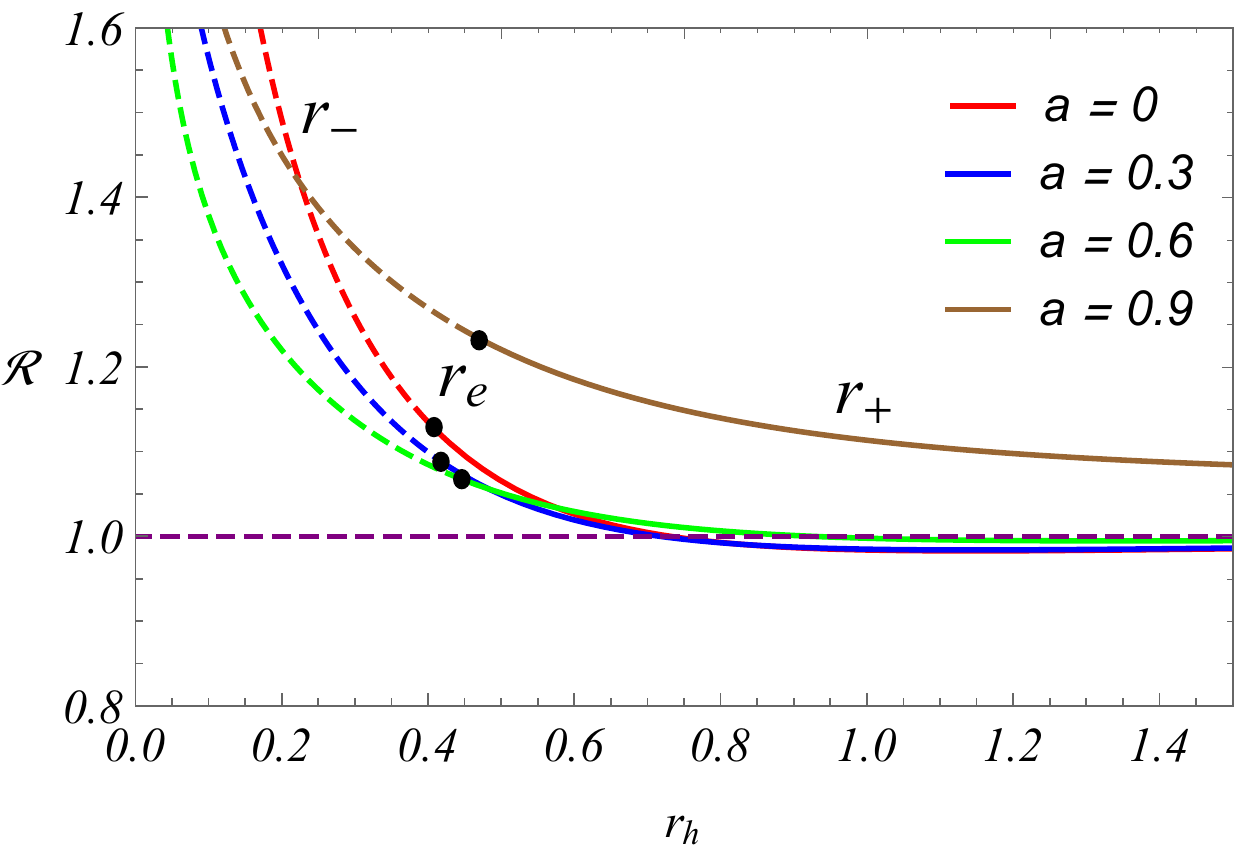}}
\caption{The isoperimetric ratio $\mathcal{R}$ vs horizon radius $r_h$ with different numerical
values of the rotation parameter $a$ and charge parameter $q$ ($l = 1$). Solid/dashed lines
correspond to outer/inner horizons, while black dots mark out the extremal horizons. Dashed
horizontal purple line corresponds to $\mathcal{R} = 1$.} \label{Rs2}
\end{figure}

As before, one can plot the mass parameter $m$ and the isoperimetric ratio $\mathcal{R}$ as
functions of horizon radius $r_h$ via
\bea
m = \frac{(r_h^2+l^2)^2(r_h^2+a^2) +qr_h(2r_h^2+l^2+a^2) +q^2}{2l^2r_h} \, , \nn
\eea
and eq. (\ref{ipro}) for different numerical values $q$ or $a$. Typically, the $m-r_h$ plot for
$q > 0$ resembles to Fig. \ref{mrq} and so will not be displayed here. Figs. \ref{Rrqa3}-\ref{Rrqa9}
exhibit the effect of the charge parameter on the isoperimetric ratio for a fixed value of the
rotation parameter; while Figs. \ref{Rraq25}-\ref{Rraq75} show the effect of the rotation parameter
on the ratio for a fixed value of the charge parameter.


Analogous to the singly-rotating case, a very crucial but subtle task is to determine where
the extremal horizons locate on the plotted curves. In the neutral case ($q = 0$), it is not
difficult to get the following expressions via solving $\Delta(r_e) = 0 = \Delta^{\prime}(r_e)$:
\be
m_e = \frac{r_e(r_e^2+l^2)^3}{l^2(l^2-3r_e^2)} \, , \qquad
r_e = \Big(\frac{\sqrt{(3a^2 +l^2)^2 +20a^2l^2} -3a^2-l^2}{10}\Big)^{1/2} \, ,
\ee
where $r_e$ is the positive root of an equation: $r_e^2(5r_e^2+l^2) +a^2(3r_e^2-l^2) = 0$. It
is easy to see that only when the extremal horizons lie in the interval ($0 < r_e < l/\sqrt{3}$)
can the the extremal mass parameter be positive: $m_e > 0$. For three chosen numerical values
of the rotation parameter $a = (0.3, 0.6, 0.9)$, their corresponding values of $r_e$, $m_e$ and
$\mathcal{R}_e$ are tabulated in Tab. \ref{exT2}, and the concrete locations of the extremal
horizons are marked out by the black dots on the red curves in Figs. \ref{Rrqa3}-\ref{Rrqa9}.

\begin{table*}[h]
\centering
\begin{tabular}{c|c|c|c}\hline
$a$ & $r_e$        & $m_e$       & $\mathcal{R}_e$ \\ \hline
0.3 & 0.2402922352 & 0.343943864 & 0.884821455 \\ \hline
0.6 & 0.3626369020 & 0.867640401 & 0.963432931 \\ \hline
0.9 & 0.4310666935 & 1.624209934 & 1.163246031 \\ \hline
\end{tabular}
\caption{The numerical values of the horizon radius $r_e$, mass parameter $m_e$, and
isoperimetric ratio $\mathcal{R}_e$ with different rotation parameter $a$ in the extremal
uncharged black hole case ($q = 0$, $l = 1$). \label{exT2}}
\end{table*}

When $q\neq 0$, there are two sets of solutions given by $\Delta(r_e) = 0 = \Delta^{\prime}(r_e)$:
\be
q = 2r_e^3 \pm\sqrt{Z_e} \, , \qquad
m_e = \frac{r_e}{l^2}\big[9r_e^4 +5l^2r_e^2 +l^4 +a^2(3r_e^2+2l^2)\big]
  \pm\frac{6r_e^2 +a^2 +l^2}{2l^2}\sqrt{Z_e} \, ,
\ee
in which $Z_e = r_e^2(3r_e^2+l^2)^2 +a^2(3r_e^2-l^2)(r_e^2+l^2) \geq 0$ must be positive to ensure
the rationality of square roots. However, we have tested actually that only the `+' sign case is
possible to avoid complex numerical results for large enough parameters $q$ and $a$, so we will
discard the `-' sign case in the numerical computations. We list the corresponding numerical values
for $r_e$, $m_e$ and $\mathcal{R}_e$ in Tab. \ref{exT3} and mark the black dots to exhibit the
concrete locations of extremal horizons on each curves in Fig. \ref{Rs2}.

\begin{table*}[h]
\centering
\begin{tabular}{l|c|c|c|c}\hline
$a$ & $q$   & $r_e$        & $m_e$       & $\mathcal{R}_e$ \\ \hline
    & 0.25  & 0.2061777056 & 0.399235187 & 1.246797393 \\ \cline{2-5}
0   & 0.5   & 0.3262967031 & 0.886060114 & 1.155596683 \\ \cline{2-5}
    & 0.75  & 0.4086893217 & 1.466753379 & 1.127194602 \\ \hline
    & 0.25  & 0.2780422204 & 0.617153563 & 1.029311697 \\ \cline{2-5}
0.3 & 0.5   & 0.3522808544 & 1.073364929 & 1.074347569 \\ \cline{2-5}
    & 0.75  & 0.4199037841 & 1.649592172 & 1.087181296 \\ \hline
    & 0.25  & 0.3672288037 & 1.156562327 & 1.006306124 \\ \cline{2-5}
0.6 & 0.5   & 0.4011346070 & 1.607131015 & 1.042716118 \\ \cline{2-5}
    & 0.75  & 0.4451693043 & 2.190443394 & 1.067775875 \\ \hline
    & 0.25  & 0.4280920102 & 1.969338567 & 1.187703259 \\ \cline{2-5}
0.9 & 0.5   & 0.4443620701 & 2.458005410 & 1.211979644 \\ \cline{2-5}
    & 0.75  & 0.4719602520 & 3.077532234 & 1.232486417 \\ \hline
\end{tabular}
\caption{The numerical values of the horizon radius $r_e$, mass parameter $m_e$, and
isoperimetric ratio $\mathcal{R}_e$ with different rotation parameter $a$ and electric
charge parameter $q$ in the extremal black hole case ($l = 1$). \label{exT3}}
\end{table*}

In Fig. \ref{Rs2}, we have plotted six figures for the isoperimetric ratio $\mathcal{R}$ as a
function of horizon radius $r_h$ for some different charge parameter $q$ and rotation parameter
$a$, and used the dashed purple line in each graph to designate the value of $\mathcal{R} = 1$.
Fig. \ref{Rrqa9} clearly shows that for larger enough rotation parameter $a$, the isoperimetric
ratio can be always larger than unity, indicating that the black hole is sub-entropic in this
situation. Each solid curve in the remaining $\mathcal{R}-r_h$ plots can intersect with and
across the dashed purple line $\mathcal{R} = 1$, which means that the sign of $(\mathcal{R} -1)$
is uncertain and in general, the doubly-rotating charged AdS$_6$ black hole when ultra-spinning
in one direction can be either sub-entropic or super-entropic in these cases.

\section{Conclusion}\label{sIV}

Taking advantage of simplicity of a solution-generating trick via taking the ultra-spinning
limit, in this paper we have constructed the ultra-spinning cousins from Chow's two equal-charge
rotating AdS black hole in six-dimensional $\cN = 4$, $SU(2)$ gauged supergravity theory.

As a first step, we have exhaustively explored the ultra-spinning black hole in the singly-rotating
case about some of its properties, namely, thermodynamics, horizon geometry, and in particular,
the RII, since all the expressions are relatively simple and very convenient for our purpose.
We have presented all necessary thermodynamical quantities and demonstrated that they obey both
the differential and integral mass formulae. Remarkably, for the six-dimensional ultra-spinning
Chow's charged black hole with only one rotation parameter, we have clearly shown that it does
not always satisfy the RII anymore, and it can be either sub-entropic or super-entropic, depending
upon the ranges of the solution parameters, especially the charge parameter. This property is
obviously different from the uncharged case --- the singly-rotating Kerr-AdS$_6$ super-entropic
black hole, which always strictly violates the RII.

We then turn to consider the doubly-rotating case, namely, the six-dimensional ultra-spinning
Chow's charged AdS black hole with an additional rotation parameter. We have explicitly shown
the effects of the charge and rotation parameters on the isoperimetric ratio and indicated that
the RII can be either violated or satisfied in the general case, and accordingly the ultra-spinning
solution can be either super-entropic or sub-entropic, depending upon on the ranges of the solution
parameters. A related subject is to extend the present work to investigate the two equal-charge
rotating black holes in five \cite{PRD72-041901} and seven \cite{CQG25-175010} dimensions. We
hope to report the related progress along this direction soon.

\acknowledgments

We thank Prof. Puxun Wu and Prof. Hongwei Yu for helpful discussions, and are greatly indebted to
the anonymous referee for invaluable comments and good suggestions to improve the presentations of
this paper. This work is supported by the National Natural Science Foundation of China (NSFC) under
Grant No. 11675130 and No. 12075084.

\end{document}